# Performance Limit of Fiber-Longitudinal Power Profile Estimation Methods

Takeo Sasai, *Member, IEEE*, Etsushi Yamazaki, *Member, IEEE,* and Yoshiaki Kisaka

(*Invited Paper*)

*Abstract*—This paper presents analytical results on power profile estimation (PPE) methods, which visualize signal power evolution in the fiber-longitudinal direction at a coherent receiver. Two types of PPE methods are reviewed and analyzed, including correlation-based methods (CMs) and minimum-mean-square-error-based methods (MMSEs). The analytical expressions for their output power profiles and spatial resolution are provided, and thus the theoretical performance limits of the two PPE methods and their differences are clarified. The derived equations indicate that the estimated power profiles of CMs can be understood as the convolution of a true power profile and a smoothing function. Consequently, the spatial resolution and measurement accuracy of CMs are limited, even under noiseless and distortionless conditions. Closed-form formulas for the spatial resolution of CMs are shown to be inversely proportional to the product of a chromatic dispersion coefficient and the square of signal bandwidth. With MMSEs, such a convolution effect is canceled out and the estimated power profiles approach a true power profile under a fine spatial step size.

## I. INTRODUCTION

CONTINUOUS monitoring of the physical parameters of transmission links is crucial to fully utilize the potential capacity of optical transmission systems during their operational lifetime and to reduce operational costs. Numerous studies have been devoted to estimating various link parameters from receiver-side (Rx) digital signal processing (DSP), such as optical signal-to-noise ratio, fiber nonlinear noise [1], [2], and total chromatic dispersion (CD) [3]. Although these optical performance monitoring solutions assist in predicting the link performance cost-effectively, these approaches estimate the cumulative parameters of the entire link and fail to monitor distributed characteristics, such as fiber loss, CD map, multiple amplifier gain spectra, and filter frequency responses. Thus, to monitor the individual link components, network operators still must rely on dedicated hardware testing devices, such as optical time-domain reflectometry (OTDR) and optical spectrum analyzers, which are costly since they are often placed in a span-by-span manner.

As another approach, novel Rx-DSP-based link monitoring solutions that reveal a signal power profile along the *fiber-longitudinal* direction without hardware devices have been proposed [4]–[10]. This longitudinal power profile estimation (PPE) has several advantages over the OTDR used in optical network monitoring: (i) multi-span characteristics are extracted simultaneously at a single coherent receiver. (ii) No probing light or additional optical components are required. (iii) it has more applications such as span-wise CD map estimation [6], [7], spatial and spectral PPE [6], [11]–[14], localization of anomaly amplifiers [6], [11]–[14], anomaly filters with a detuned center frequency [6], [15], polarization-dependent loss [16], and multi-path interference [9]. PPE, therefore, has the potential to revolutionize hardware-based approaches, unveiling the physical parameter distributions of various link components and even localizing their soft failures at a coherent receiver.

Accordingly, several PPE methods have been proposed, which can be classified into two main types: correlation-based methods (CMs) and minimum-mean-square-error-based methods (MMSEs). The original CM (the *in-situ* PPE) [4], [5] was first proposed in 2019. Subsequently, the original MMSE, which uses gradient optimization of the split-step method (SSM), was reported [6], [7]. Then several variants of CM and MMSE have been proposed, such as the Volterra-based MMSE [8], CM with nonlinear templates [9], and linear least squares [10]. These methods will be reviewed and analyzed in the following sections.

Despite the increased number of applications and methods of PPE, these PPE methods have not been theoretically examined. Consequently, the fundamental performance limits of PPE (*e.g.*, spatial resolution, SR), as well as the pros and cons of CMs and MMSEs remain unclear. Because the PPE and its applications should provide reliable and predictable monitoring performance for practical deployment, their theoretical background should be solid to minimize their uncertainty.

This study aims to develop a theoretical understanding of PPE methods for both CMs and MMSEs. Notably, analytical expressions for their estimated power profiles and SR are provided. Thus, the theoretical performance limits of PPE methods in noiseless and distortionless conditions and the differences between the two methods are clarified. A summary of our findings is as follows:

1) In CMs, the estimated power profiles are expressed in (18) and (21) under stationary Gaussian signal and constant CD assumptions. These equations indicate that the estimated power profile of CMs is a convolution of the true power profile and a spatial response function (SRF) $g(z)$. Consequently, SR is limited due to the SRF even under noiseless and distortionless conditions. Closed-form formulas for the SRs are also derived as (23) and (24), which are inversely proportional to the product of the CD coefficient and the square of the signal bandwidth.
2) In MMSEs, the estimated power profiles are expressed in (33) under the stationary Gaussian signal and constant CD assumptions. This equation implies that the convolution effect owing to the SRF in CMs is canceled out in the MMSEs. Consequently, the MMSEs approach the true power profile under a fine spatial step size.
3) CMs do not estimate the true (absolute) power owing to the SRF, whereas MMSEs do.
4) CMs have the modulation format dependency, whereas MMSEs do not.

This study extends the work presented in [17]–[19] by
- Deriving closed-form formulas for the SR of CMs.
- Adding the analysis on the spatial step size dependency of MMSEs.
- Adding discussions on the impact of noise and distortion, modulation format, and CD maps.

The remainder of this paper is organized as follows. Section II provides an overview of the system configuration of PPE methods. In Section III, analytical expressions for the power profile and SR of CMs are derived. A similar analysis for MMSEs is provided in Section IV. Section V discusses the impact of noise and distortion, modulation format, and CD maps, which are not considered in the derivations. Finally, Section VI concludes the study.

## II. OVERVIEW OF PPE

All PPE methods estimate the signal power from the nonlinear phase rotation (NLPR) at the measurement position on a fiber. The NLPR estimation problem can be regarded as an inverse problem of the nonlinear Schrödinger equation (NLSE) given in (1) and (2) [20], where the nonlinear coefficients $\gamma'(z)$ are estimated from the boundary conditions (*i.e.*, transmitted and Rx signals).

$$\frac{\partial A}{\partial z} = \left(j\frac{\beta_2(z)}{2}\frac{\partial^2}{\partial t^2} + \frac{\beta_3(z)}{2}\frac{\partial^3}{\partial t^3}\right)A - j\gamma'(z)|A|^2 A \quad (1)$$

$$\gamma'(z) \equiv \gamma(z)P(0)\exp\left(-\int_0^z \alpha(z')\,dz'\right) = \gamma(z)P(z). \quad (2)$$

Here, $A \equiv A(z,t)$, $\alpha(z)$, $\beta_2(z)$, $\beta_3(z)$, $\gamma(z)$, and $P(z)$ are the normalized optical signals at position $z$ and time $t$, fiber loss, second/third dispersion, nonlinear constant, and optical signal power at $z$, respectively. Note that $A(z,t)$ has a constant power $P_A = 1$ during propagation; in turn, all the power variations due to fiber losses and amplifications are governed only by $\gamma'(z)$ [6]. Thus, our estimation target is $\gamma'(z)$. Assuming $\gamma(z)$ is constant, the signal power profile is obtained from $\gamma'(z)$ using (2).

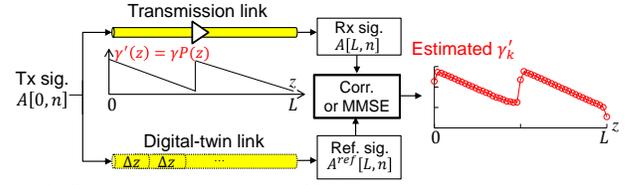
Fig. 1. Common configuration of PPE. The longitudinal power profile is obtained by conducting cross-correlation or MMSE estimation between Rx and reference signals.

A common configuration of PPE methods is shown in Fig. 1. The transmitted signals $A[0,n] = A(z=0, t=nT)$ propagate a fiber link governed by the NLSE (upper tributary) and evolve into $A[L,n]$ at a coherent receiver, where $n$ is the time sample number, $T$ is the sampling period, and $L$ is the total distance. On the other tributary, $A[0,n]$ also propagates a *digital-twin link* that emulates the fiber link in the digital domain (*e.g.*, SSM). Subsequently, the polarization rotation, frequency offset, and phase noise are aligned between the Rx signals $A[L,n]$ and digitally propagated reference signals $A^{ref}[L,n]$. For simplicity, the same notations $A[L,n]$ and $A^{ref}[L,n]$ are used for signals after this alignment. $A[L,n]$ and $A^{ref}[L,n]$ are then "compared" to estimate $\gamma'(z)$ by conducting cross-correlation or MMSE. The following sections analyze the CMs and MMSEs in detail.

*Remark 1:* In Fig. 1, forward propagation of $A[0,n]$ on the digital twin link is assumed, although most previous studies on PPE [4]–[9] used the backpropagation of Rx signals $A[L,n]$ (*i.e.*, digital backpropagation [21]). This difference is crucial when the link noise cannot be ignored because the backpropagation of Rx signals accompanies noise enhancement due to the nonlinear operation.

*Remark 2:* For simplicity, the above explanation and the following derivation assume a single polarization. The main results in this study remain unchanged if the dual-polarization transmission is analyzed based on the Manakov equation [22] instead of NLSE.

*Remark 3:* Although transmitted signals $A[0,n]$ are required for PPE, this does not mean that training or pilot signals are required because the transmitted signals can be reconstructed in the receiver through the standard demodulation process (see [4], [6]).

## III. CORRELATION-BASED METHODS

### A. Theory

Two CMs have been proposed: the original CM (the *in-situ* PPE) proposed by Tanimura *et al.* [4], [5] and the modified CM by Hahn *et al.* [9]. In this section, both methods are analyzed and the closed-form formulas for the SR are derived.

The original CM uses a simplified SSM as the digital twin link. In the simplified SSM, a partial CD operation $\widehat{D}_{0z_k}$ corresponding to the distance from the transmitter $z=0$ to a measurement point $z=z_k$ ($k=0,1,\ldots,K-1$) is first applied to $A[0,n]$. Explicitly, $\widehat{D}_{z_1 z_2} \equiv \widehat{F}^{-1}\widetilde{D}_{z_1 z_2}(\omega)\widehat{F}$, where $\widehat{F}$ is the Fourier operator, $\widetilde{D}_{z_1 z_2}(\omega) \equiv \exp\left(-j\left(\frac{\omega^2}{2}\int_{z_1}^{z_2}\beta_2(z)dz + \right.\right.$

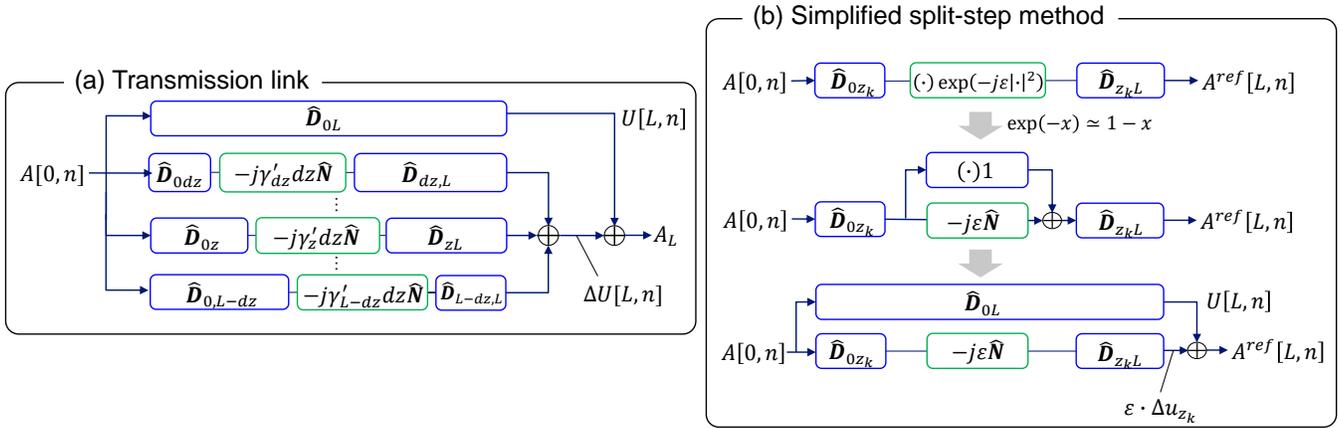

Fig. 2. First-order regular perturbation model of (a) transmission link and (b) simplified split-step method.

$\frac{\omega^3}{6} \int_{z_1}^{z_2} \beta_3(z) dz$, and $\omega$ is the angular frequency. Subsequently, a partial NLPR $\widehat{N_e}$ is applied such that

$$\widehat{N_e} \equiv (\cdot) \exp(-j\varepsilon|\cdot|^2) \quad (3)$$

where $\varepsilon$ is a fixed scaling parameter for the NLPR. The residual CD operation $\widehat{D}_{z_k L}$ from the measurement point to the link end $z = L$ is then applied. After all, $A^{ref}[L,n]$ of the original CM for the measurement point $z_k$ is expressed as follows:

$$A_k^{ref}[L,n] = \widehat{D}_{z_k L} \widehat{N_e} \widehat{D}_{0 z_k} A[0,n]. \quad (4)$$

For simplicity, the abbreviated notation $\hat{X}\hat{Y}\hat{Z}A[0,n] \equiv \hat{X}\left[\hat{Y}\left[\hat{Z}[A[0,n]]\right]\right]$ is used throughout this paper.

After $A[L,n]$ and $A_k^{ref}[L,n]$ are obtained, their cross-correlation (with no delay) is taken to obtain the estimation of $\gamma'(z)$ at $z_k$ as

$$\overline{\gamma'_{k,CM}} = \rho_0\left[A[L,n], A_k^{ref}[L,n]\right] \quad (5)$$

where $\rho_m[A[n], B[n]] = E[A[n+m]B^*[n]]$ is the cross-correlation with the time sample delay $m$ and $E[\cdot]$ is the expectation. To construct the entire longitudinal power profile, the above procedure is iterated by sweeping measurement point $z_k$ from $0 (= z_0)$ to $L (= z_{K-1})$ with an arbitrary spatial step size.

*Remark 4:* Strictly speaking, the algorithm of the CM analyzed here differs from the one used in the original papers [4], [5] in that they took the absolute values of Rx $A[L,n]$ and reference signals $A_k^{ref}[L,n]$ before the correlation while this paper does not. Taking the absolute value is not necessary solely for estimating power profiles but can remove the necessity of the frequency offset and carrier phase alignment. This paper does not take the absolute value to simplify the derivations, and clarifies the mathematical relationship among the CM, modified CM, and MMSEs. We confirmed through simulations that the main conclusion of this paper remains unchanged even if the absolute values are taken.

To derive $\overline{\gamma'_{k,CM}}$, the Rx signals $A[L,n]$ and digitally propagated reference signals $A_k^{ref}[L,n]$ are modeled using the first-order enhanced regular perturbation (eRP1) [23], [24].

First, the Rx signals with a sufficiently high sampling rate are expressed as

$$A[L,n] \simeq U[L,n] + \Delta U[L,n] \quad (6)$$

where

$$U[L,n] \equiv \widehat{D}_{0L} A[0,n] \quad (7)$$

$$\Delta U[L,n] \equiv \int_0^L \gamma'(z) \cdot \Delta u_z[L,n] dz \quad (8)$$

$$\Delta u_z[L,n] \equiv -j\widehat{D}_{zL} \widehat{N_P} \widehat{D}_{0z} A[0,n] \quad (9)$$

and

$$\widehat{N_P} \equiv (|\cdot|^2 - 2P_A)(\cdot). \quad (10)$$

The block diagram of the eRP1 approximation of Rx signals is shown in Fig. 2(a). The top path is the linear path $U[L,n]$ and the other paths $\Delta u_z[L,n]$ are partial nonlinear paths of which the summation forms $\Delta U[L,n]$.

*Remark 5:* The nonlinear operator $\widehat{N_P}$ includes the term $-2P_A(\cdot)$, which is the core idea of eRP1 that distinguishes it from RP1 [23]. RP1 uses $\widehat{N} = |\cdot|^2(\cdot)$ instead of $\widehat{N_P}$. This additional term in $\widehat{N_P}$ eliminates the phase offset incurred by the Kerr effect and is justified because the phase offset is eliminated by the carrier phase recovery in the Rx DSP and thus can be ignored in advance. This modification simplifies the calculation of correlations. In a dual-polarization case, $-2P_A$ should be replaced with $-\frac{3}{2}P_A$ [24].

Next, $A_k^{ref}[L,n]$ in (4) is also approximated using the eRP1 as shown in Fig. 2(b). The nonlinear operator is approximated as

$$\widehat{N_e}[\cdot] \simeq (\cdot) - j\varepsilon \widehat{N}[\cdot] \quad (11)$$

where $\exp(-x) \simeq 1 - x$ is used. In addition, $\widehat{N}$ can be replaced by $\widehat{N_P}$, which eliminates the phase offset in advance [24]. Consequently, (4) can be transformed as follows:

$$A_k^{ref}[L,n] = U[L,n] + \varepsilon \Delta u_{z_k}[L,n]. \quad (12)$$

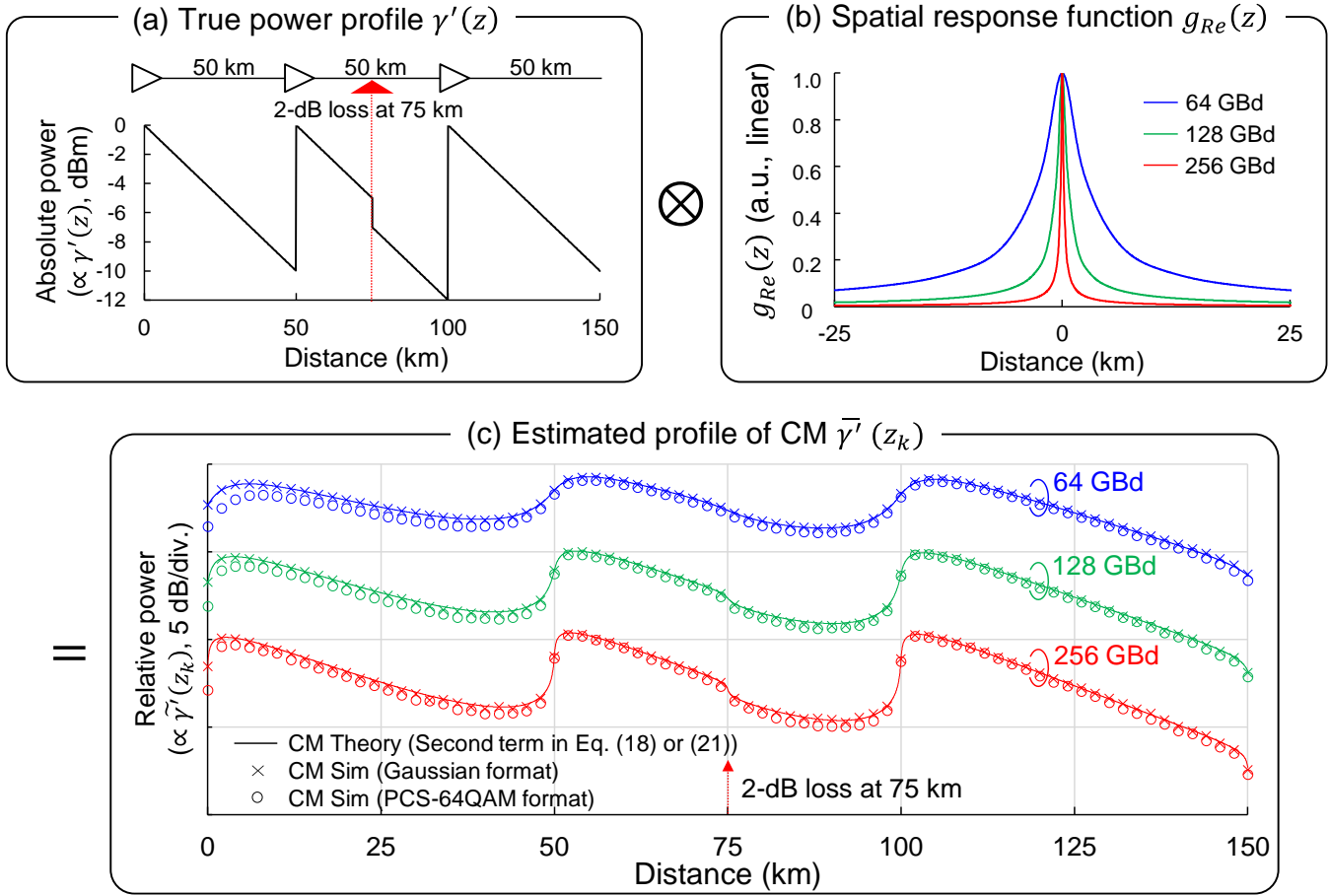

Fig. 3. Schematic of power profiles estimated using correlation-based methods (CMs). (c) The estimated power profile of CMs is a convolution of (a) true power profile and (b) spatial response function (SRF). Spatial resolution (SR) is enhanced as the symbol rate increases. In (c), power profiles predicted by (18) (solid) and simulation results (markers) are shown.

Compared with Fig. 2(a), $A_k^{ref}[L,n]$ in the original CM can be recognized as *a linear path* $U[L,n]$ + *a single nonlinear path* $\Delta u_{z_k}[L,n]$.

By substituting (6) and (12) into (5), the estimated power profile of the CM is expanded as

$$\overline{\gamma'_{k,CM}} = \rho_0[U,U] + \rho_0[U,\varepsilon\Delta u_{z_k}] \\ + \rho_0[\Delta U, U] + \rho_0[\Delta U, \varepsilon\Delta u_{z_k}]. \quad (13)$$

These correlations can be simplified under the same assumption as Gaussian noise (GN) models [24], [25], *i.e.*, $U[z,n]$ is a stationary complex Gaussian process. The derivations are described in Appendix A and are summarized as follows: The first term is equivalent to the power of $A$, *i.e.*, $P_A = 1$. The second and third terms disappear because the correlation between the linear and single nonlinear paths is zero. The fourth term is the most important term and is expressed as follows:

$$\rho_0[\Delta U, \varepsilon\Delta u_{z_k}] = 2\varepsilon \int_0^L \gamma'(z) G(z_k, z)\, dz \quad (14)$$

where

$$G(z_k, z) = \left[\widehat{D}_{z_k z}^{-1} \widehat{N} \widehat{D}_{z_k z} \rho_m[A[0,n], A[0,n]]\right]_{m=0}. \quad (15)$$

is the SRF, and $\rho_m[A[0,n], A[0,n]]$ is the autocorrelation of $A[0,n]$. Because the SRF $G(z_k,z)$ is complex-valued, the estimation of $\gamma'(z)$ is obtained by taking the real part of (13) as follows:

$$\overline{\gamma'_{k,CM}} = P_A + 2\varepsilon \int_0^L \gamma'(z)\mathrm{Re}[G(z_k,z)]\, dz. \quad (16)$$

This is the output power profile of the CM (without taking the absolute values of Rx and reference signals). Equation (16) is a Fredholm integral equation of the first kind and the SRF $G(z_k,z)$ is a kernel function, which often appear in various inverse problems. To promote understanding of (16), consider a special case in which the CD coefficients $\beta_2(z)$ and $\beta_3(z)$ are constant. Then $\widehat{D}_{z_k z} = \widehat{D}_{z_k-z,0}$ and thus $G(z_k,z) = g(z_k - z)$ holds, where

$$g(z) = \left[\widehat{D}_{z0}^{-1} \widehat{N} \widehat{D}_{z0} \rho_m[A[0,n], A[0,n]]\right]_{m=0} \quad (17)$$

is the SRF under constant CD coefficients. Equation (16) is thus reduced to a simple convolution formula:

$$\overline{\gamma'_{k,CM}} = P_A + 2\varepsilon \int_0^L \gamma'(z) g_{Re}(z_k - z)\, dz = P_A + 2\varepsilon \cdot (\gamma' \otimes_c g_{Re})(z_k) \quad (18)$$

where $\otimes_c$ denotes the continuous spatial convolution and $g_{Re}$ is the real part of $g$. This equation provides a physical understanding of the CM: *the output power profile of the CM is a filtered version of the true power profile, convolved with the SRF $g_{Re}(z)$*. This interpretation is shown schematically in Fig. 3. The details of the simulation are provided in Section III.C. Since the SRF has a low-pass filter-like characteristic (Fig. 3(b)), the output of the CM has a limited SR (Fig. 3(c)). Furthermore, CM does not estimate the true power in dBm since the SRF scales $\gamma'(z)$. In addition, (18) has an offset $P_A$, which hinders the estimation of the power change in dB (and the amount of anomaly loss). Accordingly, May *et al.* proposed a calibration method to estimate the relative power in dB [26], assuming that the power profile of the CM is an affine transformation (scaling and offset) of the true profile. This is logical from (18), which can be partly seen as an affine transformation, although it includes the convolution of the SRF.

As another countermeasure, Hahn *et al.* [9] proposed another CM (correlation with nonlinear templates) to remove the offset $P_A$, which uses $-j\widehat{N} \equiv -j|\cdot|^2(\cdot)$ instead of $\widehat{N_e}$ for the reference signal in (4). Notice that $-j\widehat{N}$ appears in the first-order term of the Taylor expansion of $\widehat{N_e}$ (11). This modification corresponds to eliminating a linear path from the reference signal at the bottom of Fig. 2(b). Furthermore, the eRP1 replaces $-j\widehat{N}$ with $-j\widehat{N_P}$, which results in the reference signal being expressed as follows:

$$A_k^{ref}[L, n] = \Delta u_{z_k}[L, n]. \quad (19)$$

Compared with $A_k^{ref}[L, n]$ for the original CM (12), one can see that the linear term $U[L, n]$ and the scaling factor $\varepsilon$ disappear in (19). Consequently, the first term in (13) and the offset $P_A$ in the output power profile of (18) disappear. The output of the modified CM is therefore expressed as follows:

$$\overline{\gamma'_{k,mCM}} = 2 \int_0^L \gamma'(z) \mathrm{Re}[G(z_k, z)]\, dz. \quad (20)$$

By assuming constant CD coefficients, this is also simplified as

$$\overline{\gamma'_{k,mCM}} = 2 \cdot (\gamma' \otimes_c g_{Re})(z_k) \quad (21)$$

*Remark 6:* What does the SRF physically mean? It represents how strongly the nonlinear path corresponding to a measurement position correlates with other nonlinear paths. In CMs, the nonlinearity (and thus the power) at $z_k$ is extracted by correlating the Rx signals with the reference signals (signals with a nonlinear operation at $z_k$). However, the nonlinear path of the reference signals has a non-zero correlation with nonlinearities from other positions. Thus, even if we attempt to extract the amount of nonlinearity at $z_k$, the resulting correlation includes nonlinearities from neighboring positions. This is a qualitative understanding of why CMs inherently have a limited SR.

*Remark 7:* The Gaussian signal assumption was made to derive analytical results. This is a convenient assumption for deriving (16) and (20) and the PPE is viable even under non-Gaussian formats such as the uniform QAM. However, as discussed in Section V.B., the output of the CMs is strongly dependent on the modulation format in shorter spans with a low accumulated CD, where the stationary Gaussian assumption is not justified. Under these conditions, (16) is invalid and requires modifications though it still helps to understand the behaviors of the CMs.

*Remark 8:* Equations (16) and (20) and their special cases for a constant CD (18) and (21) are Fredholm integral equations of the first kind, and thus a true $\gamma'(z)$ may be reconstructed from an output of the CMs by applying the inverse of $\mathrm{Re}[G(z_k, z)]$ or deconvolving $g_{Re}(z)$. There are several restrictions in doing so. This idea is strongly related to the output of the MMSEs and is discussed in detail in Remark 9.

### B. Spatial Resolution of CMs

As shown in the previous section, the power profile estimated using CMs is the convolution of a true power profile and the SRF. In this work, SR is defined as the full width at half maximum (FWHM) of the SRF. This definition corresponds to the minimal distance over which two consecutive loss (or amplification) events can be distinguished. To derive the SR, let us consider a spatial case in which the power spectrum density of the input signals $A[0, n]$ is a Gaussian function. This is because the Gaussian function is form invariant under the Fourier transform. For simplicity, $\beta_2(z) = const.$ and $\beta_3(z) = 0$ are also assumed. The extension to the Nyquist-shaped spectrum is discussed later.

Assume the input of the fiber has a power spectrum density of $\tilde{\rho}_A(\omega) \propto \exp\left(-\frac{\omega^2}{2\sigma^2}\right)$, where $\sigma^2$ is the variance. According to (17), the SRF is obtained by first applying the CD operation to the autocorrelation of the input, which can be operated in the frequency domain as $\exp\left(-\frac{\omega^2}{2\sigma^2} + j\frac{\beta_2 \omega^2 z}{2}\right)$. This is also a Gaussian function with a variance of $\sigma_1^2 = \frac{\sigma^2}{1 - j\beta_2 \sigma^2 z}$ and is transformed into the time domain as $\sqrt{\frac{\sigma_1^2}{2\pi}} \exp\left(-\frac{\sigma_1^2 (mT)^2}{2}\right)$. After being applied by the nonlinear operator $\widehat{N}$, the form remains Gaussian as $\frac{\sigma_1^2}{2\pi} \sqrt{\frac{\sigma_1^{*2}}{2\pi}} \exp\left(-\frac{(2\sigma_1^2 + \sigma_1^{*2})(mT)^2}{2}\right)$. Similarly, returning to the frequency domain and the inverse CD operation is applied. By returning to the time domain again and setting $m = 0$, a simple closed-form formula of the SRF is obtained as *the square root of a complex-valued Lorentzian function*:

$$g(z) \propto \frac{1}{\sqrt{1 + 2j\left(\frac{z}{z_{CD}}\right) + 3\left(\frac{z}{z_{CD}}\right)^2}} \quad (22)$$

where $z_{CD} = \frac{1}{|\beta_2|\sigma^2}$ is the characteristic length of the CD. Again, SR is defined as the FWHM of the SRF and thus can be obtained by solving $\frac{1}{2} = g_{Re}\left(\frac{SR}{2}\right)$, leading to $SR = \frac{1.76}{|\beta_2|\sigma^2}$. By replacing $\sigma$ with a 3-dB signal bandwidth $BW = \frac{\sqrt{2\ln 2}\sigma}{\pi}$ (Hz), the closed-form formula for SR is obtained as follows:

$$\text{SR} = \frac{0.248}{|\beta_2| \cdot \text{BW}^2} \quad (23)$$
(for Gaussian spectrum.)

Equations (22) and (23) indicate that the SRF and SR are determined by the CD coefficient and signal bandwidth. SR is enhanced as the symbol rate increases as shown in Fig. 3(b) and (c). An intuitive explanation for this observation is that the waveforms spread more rapidly under a larger CD effect, implying that the nonlinear path corresponding to a measurement position has less correlation with other nonlinear paths. Thus, a sharper SRF is observed as the symbol rate increases (Fig. 3(b)). These results are also in agreement with the observation of [4], where the accuracy of estimating the anomaly location is enhanced as the signal symbol rate increases.

However, in modern coherent optical transmission, signals are usually Nyquist-shaped to limit the signal bandwidth. Hence, we attempt to expand the Gaussian spectrum assumption to the Nyquist limit, *i.e.*, the rectangular spectrum. A comparison of the SRF calculated using (17) (rectangular spectrum input) and the closed-form (22) (Gaussian spectrum input) is shown in Fig. 4. If the two SRFs are compared under the same 3-dB bandwidth, the SRF of the Gaussian spectrum is sharper than that of the rectangular spectrum since the Gaussian spectrum has a broader frequency tail than the rectangular spectrum. Nevertheless, the shapes of the two SRFs (Gaussian case at 90.5 GHz and rectangular case at 128 GHz) do not change significantly; therefore, we infer that the SR of the rectangular spectrum is obtained by scaling that of the Gaussian spectrum. The SR for the cases with various signal bandwidths is shown in Fig. 5. The circles and squares represent the SR numerically estimated from the FWHM of the SRF using (17). As predicted, the SRs were proportional to $\frac{1}{|\beta_2|\text{BW}^2}$ in both cases. A linear fitting of the SR for the rectangular case is expressed as follows:

$$\text{SR} \simeq \frac{0.507}{|\beta_2| \cdot \text{BW}^2} \quad (24)$$
(for rectangular spectrum.)

Note that BW is equivalent to the signal symbol rate in a rectangular spectrum case. According to (24), CM achieves an SR of 6.0, 1.5, and 0.38 km with 64, 128, and 256 GBd signals, respectively, assuming a CD coefficient of a standard single-mode fiber (SSMF) $\beta_2$ = -20.55 ps²/km. Again, these values of SR are the minimal distance to distinguish two consecutive loss events.

### C. Numerical Check

In this subsection, the predicted power profile of the modified CM (21) is validated by numerical results. Two signal formats were prepared for Tx signals: a Gaussian-distributed signal and a probabilistic constellation-shaped (PCS) 64QAM with an entropy of 4.347 bits (IR = 3.305 bits, code rate = 0.826 [27]). The Nyquist roll-off factor was 0.1. The tested link was a 50-km × 3-span link emulated by the SSM with an oversampling rate of 20 samples/symbol and a spatial step size of 25 m. The fiber parameters were $\alpha$ = 0.20 dB/km, $\beta_2$ = –20.55 ps²/km, and $\gamma$ = 1.30 W⁻¹ km⁻¹. The carrier wavelength was 1555.574 nm, and the fiber launch power was 0 dBm. An intentional 2-dB attenuation was inserted at the center of the link (75 km). Amplified spontaneous emission (ASE) noise was not added to investigate the performance limit. No CD predistortion was applied to the Tx signals.

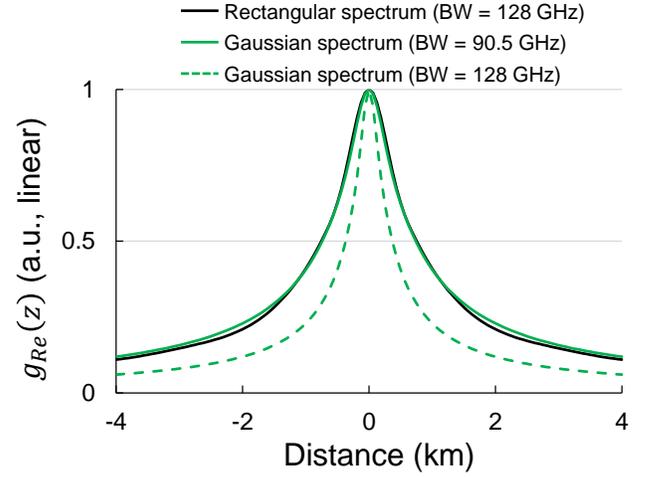

Fig. 4. Comparison of spatial response functions (SRFs) for rectangular and Gaussian input spectra. BW denotes 3-dB bandwidth of spectrum. $\beta_2$ = -20.55 ps²/km is assumed.

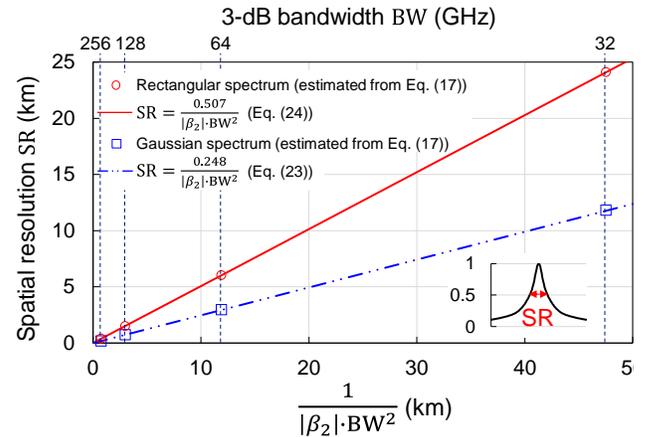

Fig. 5. Spatial resolution of CMs for rectangular and Gaussian spectra with various 3-dB bandwidth. $\beta_2$ = -20.55 ps²/km is assumed.

The power profiles predicted using the second term in (18) and its equivalent (21) (solid) and those obtained through simulations (cross and circle) are shown in Fig. 3(c). The spatial granularity $\Delta z$ for PPE was set to 2 km. The theoretical lines are in good agreement with the simulation results. Particularly, the simulation results of the Gaussian distribution format (cross) agreed with the theoretical results, whereas the results of the PCS-64QAM (circle) deviated slightly near the Tx side. This is because (21) is derived assuming that $U[z, n]$ is a stationary complex Gaussian process. The PCS-64QAM format does not satisfy this assumption in the low CD area, *i.e.*, Tx side; however, the assumption is gradually satisfied as the distance increases due to the accumulated CD, and thus the deviation diminishes. The dependency on other modulation formats is discussed in Section V.B. As described in Section III.A. and III.B., the SR of the CMs is limited by the convolution effect owing to the SRF $g_{Re}(z)$, even under noiseless and distortionless conditions. However, as the baud rate increases, $g_{Re}(z)$ becomes more like a delta function due to a more chromatic dispersion effect (Fig. 3(b)), resulting in sharper power profiles and an enhanced SR.

## IV. MMSE-BASED METHODS

### A. Theory

In MMSEs, $\overline{\gamma'_k}$ is estimated as the optimum nonlinear coefficient in a digital twin link that best emulates the actual fiber link. This is formulated as a classical least-squares problem:

$$\overline{\gamma'_k} = \underset{\gamma'_k}{\arg\min} I = \underset{\gamma'_k}{\arg\min} E\left[\left|A[L,n] - A^{ref}[L,n]\right|^2\right]. \quad (25)$$

Note that $A^{ref}[L,n]$ is a function of $\gamma'_k$.

The original MMSE was proposed in [6], [7] and uses the SSM as the digital twin link to obtain $A^{ref}[L,n]$. The nonlinear coefficients $\gamma'_k$ in the SSM are optimized using gradient descent to minimize the cost function $I$. Gleb *et al.* [8] proposed the Volterra-series expansion to model $A^{ref}[L,n]$. A linear least squares method for PPE [10] was also proposed, where $A^{ref}[L,n]$ was modeled using RP1. MMSEs differs from CMs in that they use a complete model of transmission links (*e.g.*, a full SSM) as a digital twin link rather than a simplified SSM.

Equation (25) is a nonlinear inverse problem of the NLSE, where the parameters in the equation are estimated from the boundary conditions, *i.e.*, the Tx and Rx signals. Such inverse problems are often ill-posed: many local minima may exist. However, the uniqueness of the solution can be guaranteed under the RP1 approximation and a monotonically increasing CD (*i.e.*, dispersion unmanaged link, see Remark 10 and Section V.C.), which enables deriving an analytical expression of $\gamma'_k$.

$\overline{\gamma'_k}$ is derived by approximating both $A[L,n]$ and $A^{ref}[L,n]$ using eRP1 and substituting them into cost function $I$ in (25). $A[L,n]$ was previously described in (6). $A^{ref}[L,n]$ is expressed as a discretized version of $A[L,n]$. The cost function is then transformed as follows:

$$I \simeq E\left[\left|\Delta U[L,n] - \Delta U^{ref}[L,n]\right|^2\right] \quad (26)$$

where $U[L,n] = U^{ref}[L,n]$ is used, assuming the sampling rate satisfies the Nyquist theorem for the linear terms. Notice that the cost function becomes a quadratic function of $\gamma'_k$ because $\Delta U^{ref}[L,n]$ is a linear function of $\gamma'_k$. Thus, the nonlinear least squares problem (25) is now reduced to a linear least squares problem, which provides an analytical expression for $\overline{\gamma'_k}$. By expanding the expectation, we obtain

$$I \simeq \rho_0[\Delta U, \Delta U] + \rho_0[\Delta U^{ref}, \Delta U^{ref}] - 2\text{Re}\left[\rho_0[\Delta U, \Delta U^{ref}]\right]. \quad (27)$$

The derivation of the partial derivative $\partial I/\partial \gamma'_k$ is described in Appendix B and summarized as follows: The first term disappears when calculating $\partial I/\partial \gamma'_k$ since $\Delta U$ is independent of $\gamma'_k$. The second and third terms can be calculated similarly as in (13) under the stationary Gaussian signal assumption. The partial derivative is then expressed as follows:

$$\frac{\partial I}{\partial \gamma'_k} = 4\Delta z^2 \sum_{i=0}^{K-1} \gamma'_i \text{Re}[G(z_k, z_i)] - 4\Delta z \int_0^L \gamma'(z) \text{Re}[G(z_k, z)] \, dz. \quad (28)$$

By solving $\partial I/\partial \gamma'_k = 0$, we obtain

$$\sum_{i=0}^{K-1} \overline{\gamma'_i} \text{Re}[G(z_k, z_i)] \Delta z = \int_0^L \gamma'(z) \text{Re}[G(z_k, z)] \, dz. \quad (29)$$

$\overline{\gamma'_k}$ can be obtained by applying the inverse of $\text{Re}[G(z_k, z_i)]$ to both sides in a matrix form as described in [10]. In this paper, to promote an understanding of MMSEs and their relation to CMs, let us again assume a constant CD as in the derivation of (18). Equation (29) is then transformed as

$$\sum_{i=0}^{K-1} \overline{\gamma'_i} g_{Re}(z_k - z_i) \Delta z = \int_0^L \gamma'(z) g_{Re}(z_k - z) \, dz. \quad (30)$$

or simply

$$(\overline{\gamma'} \otimes_d g_{Re})(z_k) \cdot \Delta z = (\gamma' \otimes_c g_{Re})(z_k) \quad (31)$$

where $\otimes_d$ denotes the discrete spatial convolution. Notice that the right-hand side of (31) is the same as the output of the modified CM (21) except for a scaling factor of 2. This *discrete vs. continuous* convolution equation provides an intuitive understanding of the SR of MMSEs. As the spatial step size decreases, the discrete convolution in (30) approaches a continuous convolution on the right-hand side, implying $\overline{\gamma'_k}$ approaches the true $\gamma'(z)$. Thus, the SR of MMSE is not as limited by $g(z)$ as that of CM, and the MMSE can estimate the true power. To obtain the expression of the output power profiles of the MMSEs, the discrete Fourier transform with respect to $k$ is applied to (31), converting from the spatial domain to the wave number domain as

$$\hat{F}[\overline{\gamma'_k}] \cdot \hat{F}[g_{Re}(z_k)] \cdot \Delta z = \hat{F}[(\gamma' \otimes_c g_{Re})(z_k)]. \quad (32)$$

By deconvolving the SRF $g_{Re}(z_k)$ from both sides and returning to the spatial domain, $\overline{\gamma'_k}$ is obtained as

$$\overline{\gamma'_{k,MMSE}} = \frac{1}{\Delta z} \hat{F}^{-1}\left[\frac{\hat{F}[(\gamma' \otimes_c g_{Re})(z_k)]}{\hat{F}[g_{Re}(z_k)]}\right]. \quad (33)$$

This is the output of the MMSEs. Equation (33) implies that the convolution effect embedded in the CMs is naturally canceled out and thus a high SR is achieved in the MMSEs. The matrix form of (33) can be found in [10].

*Remark 9*: From (33), one may argue that the output power profile of the modified CM (21) can also be improved to that of MMSEs by deconvolving the SRF $g_{Re}(z_k)$. This is true if (i) transmitted signals satisfy the stationary Gaussian assumption,

(ii) the output of the CM for out of the link $z_k < 0$ and $z_k > L$ is available, and (iii) a CD map is uniform. Regarding (i), (21) and its generalization (20) were derived under the stationary Gaussian signal assumption and are invalid in the non-Gaussian area such as the shorter spans with QPSK, implying that the deconvolution is not justified. As shown in Section V.B., the power profile of CMs is strongly dependent on the modulation format in shorter spans while that of MMSEs does not. Regarding (ii), the SRF has an infinite impulse response (see (22) and Fig. 3(b)) and thus an output of the CMs has non-zero values for $z_k < 0$ and $z_k > L$. To fully deconvolve the SRF, a sufficiently long power profile ranging out of the link is required, which increases the computational complexity. Regarding (iii), under non-uniform CD maps, the generalized SRF $G(z_k, z)$ in (20) should be used instead of (21), which means a simple deconvolution of the SRF $g_{Re}(z_k)$ is not appropriate. This is not a big problem, however, because one can apply the inverse operation of $Re[G(z_k, z)]$ to the output of the modified CMs instead of the deconvolution.

*Remark 10*: The inverse of $\text{Re}[G(z_k, z_i)]$ should exist to solve (29) for $\overline{\gamma'_k}$. However, in dispersion managed systems where the fibers with opposite-sign CD coefficients exist, this inverse does not exist and thus the MMSEs fail. A monotonically increasing or decreasing CD (*i.e.*, $\beta_2(z) > 0$ or $\beta_2(z) < 0$ for all $z$) is at least required for the inverse to exist. Similarly, the CMs also do not work in dispersion-managed systems. The intuitive understanding of this reason is discussed in Section V.C.

### B. Numerical Check

First, the effectiveness of (33) is examined. The signal was 128-GBd PCS-64QAM, and the other simulation parameters were set to the same condition as in Section III.C. The theoretical (dashed line) and simulated (circle) power profiles of the MMSEs with different $\Delta z$ are shown in Fig. 6. The linear least squares method [10] was used for the simulation results. The true power profile (solid black line) is also shown for reference. As previously mentioned, noise and distortion were not added to investigate the performance limit. The theoretical line (33) accurately predicted the simulation results in all cases. The power profiles of MMSEs deviate from the true power under coarse $\Delta z$ but the deviations diminish as $\Delta z$ approaches zero and the power profiles show good measurement accuracy and SR. In particular, the loss event at 75 km is detected and its amount of 2 dB can be reliably estimated when $\Delta z$ = 1 km, implying that the MMSEs can function as an alternative to OTDR at its upper-performance limit.

Next, the power profiles of the CMs and MMSEs were obtained under ASE noise to check the applicability of the derived results to more practical conditions (Fig. 7). The noise figures of the amplifiers were set to 5.0 dB, and the symbol rate was fixed at 128 GBd. $\Delta z$ was set to 0.5 km but decimated to 2 km when plotting profiles. The number of samples used for the correlation and MMSE was fixed at $2^{20}$. The modified CM (21) was used for CM. Note that the second vertical axis was used for the CM profiles because the CM does not estimate the true power. The power profile of the MMSE accurately agrees with the true power profile, meaning the true link parameters, such as the fiber loss coefficients, fiber launch power, and the

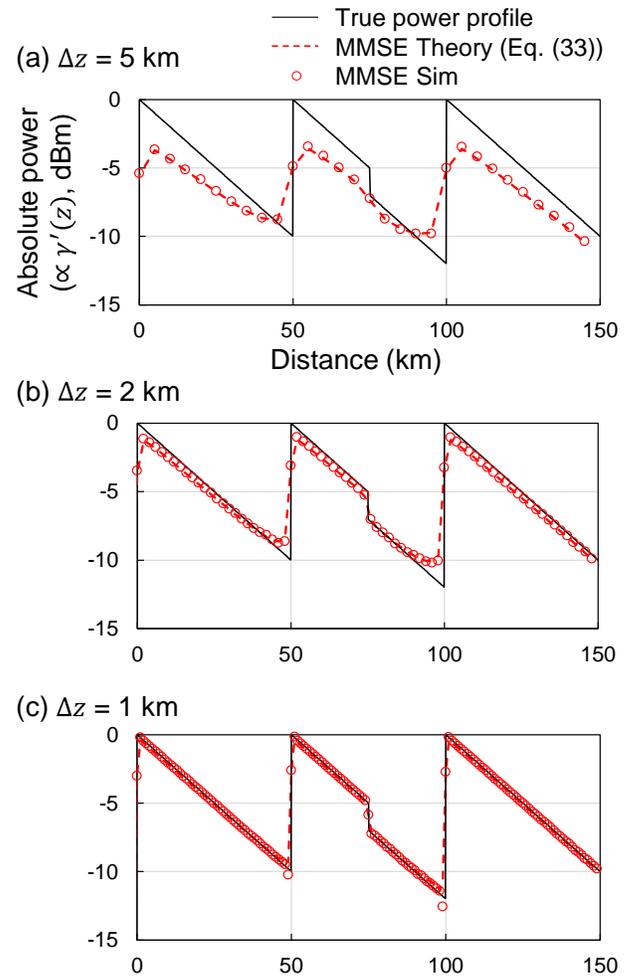

Fig. 6. Power profiles estimated using MMSE-based methods. The estimated power profile approaches the true power profile under a fine spatial step size.

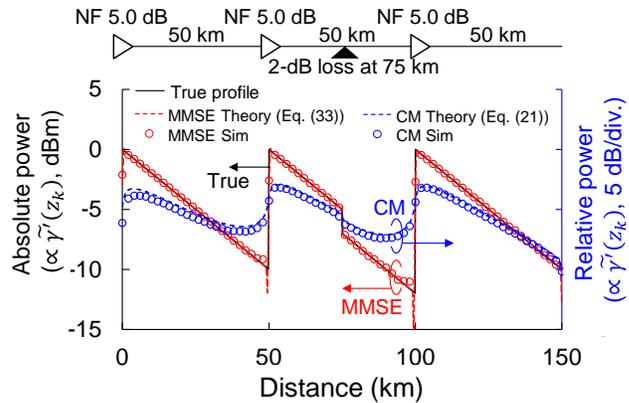

Fig. 7. Comparison of power profiles of CMs and MMSE-based methods. The second vertical axis is used for CM since CM does not estimate the true power. The symbol rate was 128 GBd.

position and amount of anomaly loss can be accurately estimated. However, the estimated power profile of the CM

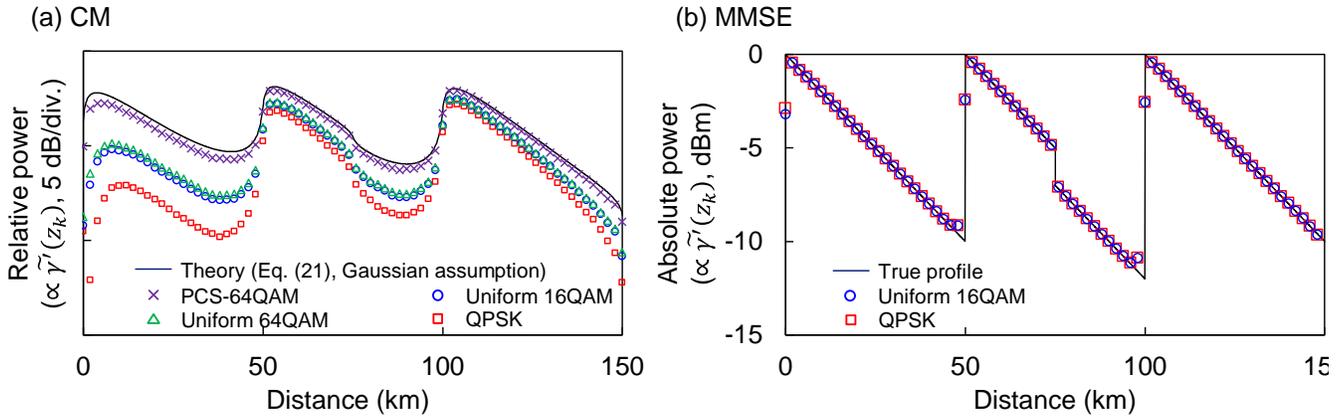

Fig. 8. Modulation format dependency of (a) modified CM and (b) MMSEs. In (b), other modulation formats such as uniform 64QAM, PCS-64QAM, and Gaussian format are not shown for visibility but reproduced same power profiles. The symbol rate was 128 GBd. No noise was added.

deviates significantly from the true power profile, and the direct estimation of such link parameters is challenging. Thus, the normal state reference, *i.e.*, the power profile without anomalies, is required to identify the position of the loss event. Moreover, a calibration method, such as one proposed by May *et al.* [26], is indispensable for correctly estimating the amount of loss in dB. These limitations of CM originate from the convolution effect of the SRF $g(z)$.

## V. UNCONSIDERED EFFECT

### A. Noise and Distortion

The derivations in this paper did not consider noise and distortion. Qualitatively speaking, the accuracy of the PPE at the measurement position $z_k$ in the presence of noise and distortion is (at least partly) dependent on the following nonlinearity-to-noise ratio (NLNR) [6]:

$$\text{NLNR}(z_k) = \frac{P_{SCI}(z = z_k)}{P_N + P_D}. \quad (34)$$

This is an analogy of the signal-to-noise ratio (SNR) in communication theory. $P_{SCI}(z = z_k)$ is the power of the self-channel nonlinear interference (SCI) at measurement position $z_k$, $P_N$ is the power of stochastic noise such as ASE noise and laser phase noise, and $P_D$ is the power of static distortion including transceiver imperfection and in-line optical filtering.

Regarding $P_{SCI}$, a high fiber launch power of channel under test (CUT) is desirable to achieve adequate measurement accuracy as the PPE relies on the SCI; otherwise, the output power profiles become noisy. However, the optimal operational launch power for communication purposes is typically not sufficient for a stable PPE performance. Especially in the latter half of spans where the signal power decreases, the measurement accuracy significantly degrades because of insufficient nonlinearity. Experimental demonstrations thus far [4]-[16] used a 4- to 10-dBm launch power per channel for 64-GBd-class signals, which is typically higher than the optimal operational power. Such a high launch power affects the SNR of other WDM channels due to the excessive nonlinearity and reducing $P_N$ and $P_D$ is therefore essential.

The impact of stochastic noise $P_N$ can be removed by increasing the number of samples used in the correlation and MMSE due to the averaging effect. However, when the PPE is used for real-time network monitoring, the number of processed samples may be limited for faster processing and lower power consumption. For such an application, residual noise disturbs the power profiles and further analysis is required to fully understand the behavior of the PPE under stochastic noise and the required number of samples for a stable estimation.

The impact of static distortion $P_D$ can be reduced by equalizing distortion. Transceiver imperfections including frequency responses, IQ imbalances, and cross-talks should be aligned between Rx signals and reference signals. These distortion contaminate a power profile in different ways and further studies are required. At least, the bandwidth limitation will limit the SR of the PPE as discussed in Section III.B. Similarly, optical filters at reconfigurable optical add/drop multiplexer (ROADM) nodes will also limit the SR.

Noise and distortion have different impacts on the CMs and MMSEs: the CMs are more robust to noise and distortion than the MMSEs. This is evident from (33), indicating that the output of the MMSEs is considered a deconvolved version of the CMs. The SRF is a low-pass filter and thus its deconvolution behaves as a high-pass filter, which enhances the noise and distortion.

### B. Modulation Format

The derivations in this paper assumed that the signals were a stationary complex Gaussian process for all $z$. However, this assumption is not justified when practical formats such as uniform QAM are used. In such a case, the output power profiles of the CMs cannot be expressed in a simple form as in (16) and (20). Fig. 8 shows the modulation formant dependency of the CMs and MMSEs. Again, the symbol rate was fixed at 128 GBd, and the other simulation conditions were the same as in Section III.C. $\Delta z$ was set to 0.5 km but decimated to 2 km when plotting profiles. Although only QPSK and uniform 16QAM cases are shown in the MMSEs for visibility (Fig. 8(b)), we confirmed that other modulation formats reproduced the same power profiles.

In CMs (Fig. 8(a)), the power profiles near the Tx side are strongly dependent on the modulation format, which means the relative power level diagram is difficult to estimate in shorter spans. This is because the signal distribution is not Gaussian under a low accumulated CD. For a more accurate prediction, the analyses in the enhanced GN model [28], [29] may be helpful to include the modulation format dependency. A countermeasure to eliminate the modulation format dependency is to apply the CD predistortion to the Tx signals as used in [4], [5], which "Gaussianizes" the Tx signals in advance. However, when data-carrying signals are used for PPE, the CD predistortion increases computational complexity in the Tx and even Rx DSP for its compensation for communication purposes. If arbitrary signal formats (*i.e.*, non-data-carrying signals) are allowed for PPE, the use of a Gaussian format will suffice rather than applying the CD predistortion.

In MMSEs, the power profile is independent of the modulation format as shown in Fig. 8(b). Though the true reason requires further analysis, this is because the SRF effect on the right-hand side of (29) or (31) is canceled out (deconvolved) by the SRF itself, and thus the modulation format dependency in the SRF is also canceled out.

### C. Chromatic Dispersion Map

A non-uniform CD map significantly affects the performance of the PPE (especially CMs) since the spatial resolution is dependent on CD coefficients as in (23) and (24). The output power profile for non-uniform CD maps should be discussed using a general form of the SRF $G(z_k, z)$ as in (16) and (20), instead of $g(z)$ in (18) and (20). This generalized SRF $G(z_k, z)$ is dependent on the measurement position $z_k$, and thus the spatial resolution also differs with $z_k$. For example, when SSMFs and non-zero dispersion shifted fibers (NZDSF) coexist in a link, the spatial resolution will be lower in the NZDSF spans. Furthermore, in a dispersion-managed link where the fibers with opposite-sign CD coefficients exist, the PPE (both the CMs and MMSEs) fails. This is because multiple positions in a link share the same amount of the accumulated CD, which implies two nonlinear paths for different positions in Fig. 2(a) have the completely same operations CD $\widehat{D}_{0z}$, nonlinearity $\widehat{N}$, and residual CD $\widehat{D}_{zL}$. The PPE cannot distinguish these two paths and becomes an ill-posed problem. Therefore, an accumulated CD should at least be monotonically increasing or decreasing (*i.e.*, $\beta_2(z) > 0$ or $\beta_2(z) < 0$ for all $z$) for a meaningful PPE. This is a current limitation of the PPE.

## VI. CONCLUSION

This study presented analytical expressions for Rx-DSP-based longitudinal PPE methods including CMs and MMSEs. First-order regular perturbation theory was applied to model the Rx and reference signals to derive the estimated power profiles and SR. The derived equations were validated by comparing them with simulation results and showed good agreement.

The derived equations for CMs (18) and (21) indicate that the estimated power profiles are the convolution of a true power profile and an SRF, resulting in limited SR and measurement accuracy. Assuming the input signal spectrum is a Gaussian function, the SRF is expressed in a simple closed form as a square root of the Lorentzian function (22). Accordingly, a closed-form formula for the SR of the Gaussian spectra (23) was obtained and extended to the Nyquist-limit (rectangular) spectra (24).

According to (33), in MMSEs, the convolution effect of the SRF that accompanied the power profiles of CMs is naturally deconvolved. Thus, the estimated power profile of the MMSE approaches the true power profile under $\Delta z \to 0$.

Extending to the case under non-negligible ASE noise, transceiver impairments, and cross-channel nonlinear interference requires further studies and constitutes scope for future research.

## APPENDIX A

### DERIVATION OF (16)

This appendix calculates the correlations in (13) to obtain (16). For simplicity, the simplified notations such as $A_0 = A[0, n]$, $U_z = U_z[n] = U[z, n]$, $\Delta U_z = \Delta U[z, n]$, and $\Delta u_{z_k, L} = \Delta u_{z_k}[L, n]$, are used.

First, we prove that the correlation between the linear path and a single nonlinear path in Fig. 2, $\rho_0[U_L, \varepsilon \Delta u_{z_k, L}] = \varepsilon \rho_0[U_L, \Delta u_{z_k, L}]$ (the second term in (13)), is zero.

$$\rho_m[U_L, \Delta u_{z_k, L}]$$
$$= \rho_m[U_L, -j\widehat{D}_{z_k L}\widehat{N}_P \widehat{D}_{0z_k} A_0]$$
$$= -j\rho_m\left[U_L, \widehat{D}_{z_k L}|U_{z_k}|^2 U_{z_k} - 2P_A U_L\right]$$
$$= -j\rho_m\left[U_L, \widehat{D}_{z_k L}|U_{z_k}|^2 U_{z_k}\right] + 2jP_A \rho_m[U_L, U_L]. \quad (35)$$

Using (59) and (60) in Appendix C, $\widehat{D}_{z_k L}$ in the first term and $\widehat{D}_{0L}$ in the second term in (35) disappear as follows:

$$\rho_m[U_L, \Delta u_{z_k, L}]$$
$$= -j\rho_m\left[U_{z_k}, |U_{z_k}|^2 U_{z_k}\right] + 2jP_A \rho_m[A_0, A_0] \quad (36)$$

where $\rho_m[A_0[n], A_0[n]]$ are autocorrelations of the transmitted signals. The first correlation in (36) can be simplified using the moment theorem (62) in Appendix D as follows:

$$\rho_m\left[U_{z_k}, |U_{z_k}|^2 U_{z_k}\right]$$
$$= E\left[U_{z_k}[n+m]|U_{z_k}[n]|^2 U_{z_k}^*[n]\right]$$
$$= 2E\left[U_{z_k}[n+m]U_{z_k}^*[n]\right]E\left[|U_{z_k}[n]|^2\right]$$
$$= 2P_A \rho_m[A_0, A_0] \quad (37)$$

where $E\left[|U_{z_k}[n]|^2\right] = P_A$ was used. By substituting (37) into (36), it becomes zero.

The third term in (13) is $\rho_0[\Delta U_L, U_L]$; however, this term is also zero because $\Delta U_L$ is the integral of $\Delta u_{z,L}$ with respect to $z$.

Finally, we calculate the last term in (13), $\rho_0[\Delta U_L, \varepsilon \Delta u_{z_k, L}] = \varepsilon \rho_0[\Delta U_L, \Delta u_{z_k, L}]$. First,

$$\rho_m[\Delta U_L, \Delta u_{z_k, L}]$$
$$= E[\Delta U_L[n+m]\Delta u_{z_k, L}^*[n]]$$
$$= \int_0^L \gamma'(z) \cdot E\big[(\widehat{D}_{zL} N_z[n+m])$$
$$\qquad \cdot (\widehat{D}_{z_k L} N_{z_k}[n])^*\big] dz$$
$$= \int_0^L \gamma'(z) \cdot \rho_m[\widehat{D}_{zL} N_z, \widehat{D}_{z_k L} N_{z_k}] dz \quad (38)$$

where

$$N_z[n] \equiv \widehat{N_P} U_z[n] = \widehat{N_P}\widehat{D}_{0z}A_0[n]. \quad (39)$$

Using (59) in Appendix C, the CD operations are taken out of the correlation as follows:

$$\rho_m[\widehat{D}_{zL}N_z, \widehat{D}_{z_kL}N_{z_k}] = \widehat{D}_{zL}\widehat{D}_{z_kL}^{-1}\rho_m[N_z, N_{z_k}]$$
$$= \widehat{D}_{zz_k}\rho_m[N_z, N_{z_k}]. \quad (40)$$

The correlation in (40) is expanded as follows:

$$\rho_m[N_z, N_{z_k}] = \rho_m\left[|U_z|^2 U_z, |U_{z_k}|^2 U_{z_k}\right]$$
$$- 2P_A \rho_m[|U_z|^2 U_z, U_{z_k}]$$
$$- 2P_A \rho_m\left[U_z, |U_{z_k}|^2 U_{z_k}\right] \quad (41)$$
$$+ 4P_A^2 \rho_m[U_z, U_{z_k}].$$

Again, the moment theorem in Appendix D simplifies these correlations. The first term of (41) is

$$\rho_m\left[|U_z|^2 U_z, |U_{z_k}|^2 U_{z_k}\right]$$
$$= E\left[|U_z[n+m]|^2 U_z[n+m]|U_{z_k}[n]|^2 U_{z_k}^*[n]\right]$$
$$= 4E[|U_z[n+m]|^2]E\left[U_z[n+m]U_{z_k}^*[n]\right]E\left[|U_{z_k}[n]|^2\right]$$
$$+ 2E\left[U_z[n+m]U_{z_k}^*[n]\right]^2 E\left[U_z^*[n+m]U_{z_k}[n]\right]$$
$$= 4P_A^2 R_{z_kz}[m] + 2|R_{z_kz}[m]|^2 R_{z_kz}[m] \quad (42)$$

where

$$R_{z_kz}[m] = E\left[U_z[n+m]U_{z_k}^*[n]\right]$$
$$= \widehat{D}_{0z}\widehat{D}_{0z_k}^{-1}\rho_m[A_0, A_0]$$
$$= \widehat{D}_{z_kz}\rho_m[A_0, A_0]. \quad (43)$$

Similarly, the second and third terms of (41) are

$$\rho_m[|U_z|^2 U_z, U_{z_k}] = \rho_m\left[U_z, |U_{z_k}|^2 U_{z_k}\right]$$
$$= 2P_A R_{z_kz}[m]. \quad (44)$$

By substituting (42) and (44) into (41), the terms related to $P_A$ disappear, and we obtain

$$\rho_m[N_z, N_{z_k}] = 2|R_{z_kz}[m]|^2 R_{z_kz}[m]$$
$$= 2\widehat{N} R_{z_kz}[m] \quad (45)$$
$$= 2\widehat{N}\widehat{D}_{z_kz}\rho_m[A_0, A_0].$$

From the second to the third line, (43) was used. Thus, (38) becomes

$$\rho_m[\Delta U_L, \Delta u_{z_k,L}]$$
$$= 2\int_0^L \gamma'(z) \cdot \widehat{D}_{z_kz}^{-1}\widehat{N}\widehat{D}_{z_kz}\rho_m[A_0, A_0]\, dz \quad (46)$$

where $\widehat{D}_{zz_k} = \widehat{D}_{z_kz}^{-1}$ is used. By substituting $m=0$, we finally obtain (16).

## APPENDIX B

### DERIVATION OF (28)

This appendix derives the partial derivative $\partial I/\partial \gamma_k'$ in (28) from (27). The partial derivative of (27) is expressed as

$$\frac{\partial I}{\partial \gamma_k'} = \frac{\partial}{\partial \gamma_k'}\left[\rho_0[\Delta U^{ref}, \Delta U^{ref}]\right]$$
$$- 2\mathrm{Re}\left[\frac{\partial}{\partial \gamma_k'}\left[\rho_0[\Delta U, \Delta U^{ref}]\right]\right] \quad (47)$$

where $\Delta U$ is expressed in (8), $\Delta U^{ref}$ is a discretized version of (8) such that

$$\Delta U^{ref}[L, n] \equiv \sum_{i=0}^{K-1} \gamma_i' \, \Delta u_{z_i}[L, n]\Delta z \quad (48)$$

and $\Delta u_{z_i}$ is defined in (9). The first term of (47) is

$$\frac{\partial}{\partial \gamma_k'}\left[\rho_0[\Delta U^{ref}, \Delta U^{ref}]\right]$$
$$= \frac{\partial}{\partial \gamma_k'}E\left[\left|\Delta U_L^{ref}\right|^2\right] \quad (49)$$
$$= 2\mathrm{Re}\left[E\left[\Delta U^{ref}\frac{\partial \Delta U^{ref*}}{\partial \gamma_k'}\right]\right].$$

Here, from (48),

$$\frac{\partial \Delta U^{ref*}}{\partial \gamma_k'} = \Delta u_{z_k}^*[L, n]. \quad (50)$$

Thus,

$$\frac{\partial}{\partial \gamma_k'}\left[\rho_0[\Delta U^{ref}, \Delta U^{ref}]\right]$$
$$= 2\mathrm{Re}\left[E[\Delta U^{ref}\Delta u_{z_k}^*]\right]. \quad (51)$$
$$= 2\mathrm{Re}\left[\rho_0[\Delta U^{ref}, \Delta u_{z_k}]\right]$$

The second term of (47) is calculated in the same way and thus (47) becomes

$$\frac{\partial I}{\partial \gamma_k'} = 2\mathrm{Re}\left[\rho_0[\Delta U^{ref}, \Delta u_{z_k}]\right]$$
$$- 2\mathrm{Re}\left[\rho_0[\Delta U, \Delta u_{z_k}]\right] \quad (52)$$

$\rho_0[\Delta U, \Delta u_{z_k}]$ appeared in the derivation of CMs and was already calculated in (14). $\rho_0[\Delta U^{ref}, \Delta u_{z_k}]$ is its discretized version, expressed as

$$\rho_0[\Delta U^{ref}, \Delta u_{z_k}] = 2(\Delta z)^2 \sum_{i=0}^{K-1} \gamma_i' \, G(z_k, z_i) \quad (53)$$

Thus, we obtain (28).

## APPENDIX C

### CORRELATION BETWEEN OUTPUTS OF LTI SYSTEMS

The correlation between the output signals of linear time-invariant (LTI) systems can be related to the correlation of their inputs [24], [30]. Let $A[n]$ and $B[n]$ be stationary random processes, and

$$C[n] = h[n] \otimes A[n] \quad (54)$$

$$D[n] = g[n] \otimes B[n]. \tag{55}$$

The cross-correlation between $C$ and $D$ is then

$$\rho_m[C, D] = (h[m] \otimes g^*[-m]) \otimes \rho_m[A, B]. \tag{56}$$

Taking the Fourier transform, we have

$$\hat{F}\rho_m[C, D] = \tilde{h}(\omega) \cdot \tilde{g}^*(\omega) \cdot \tilde{\rho}_{AB}(\omega). \tag{57}$$

Since the CD operation satisfies $\tilde{D}^*_{zz'}(\omega) = \tilde{D}^{-1}_{zz'}(\omega)$, we have

$$\hat{F}\rho_m\big[\hat{D}_{z_1z_2}A, \hat{D}_{z_3z_4}B\big] = \tilde{D}_{z_1z_2}(\omega) \cdot \tilde{D}^{-1}_{z_3z_4}(\omega) \cdot \tilde{\rho}_{AB}(\omega). \tag{58}$$

Thus, we obtain

$$\rho_m\big[\hat{D}_{z_1z_2}A, \hat{D}_{z_3z_4}B\big] = \hat{D}_{z_1z_2}\hat{D}^{-1}_{z_3z_4}\rho_m[A, B]. \tag{59}$$

In the special case of $\hat{D}_{z_1z_2} = \hat{D}_{z_3z_4}$, the CD operations cancel out:

$$\rho_m\big[\hat{D}_{z_1z_2}A, \hat{D}_{z_1z_2}B\big] = \rho_m[A, B]. \tag{60}$$

## APPENDIX D

### MOMENT THEOREM FOR COMPLEX GAUSSIAN PROCESSES

According to the moment theorem [31], zero-mean complex Gaussian random processes $U_1, \ldots, U_k, V_1, \ldots, V_k$ satisfy

$$\begin{aligned}&E[U_1 U_2 \ldots U_k V_1^* V_2^* \ldots V_k^*] \\ &= \sum_\pi E[U_1 V_{\pi_1}] E[U_2 V_{\pi_2}] \ldots [U_k V_{\pi_k}]\end{aligned} \tag{61}$$

where $\pi$ is a permutation of the set of integers $\{1, 2, \ldots, k\}$. For instance, when $k = 2$,

$$\begin{aligned}&E[U_1 U_2 V_1^* V_2^*] \\ &= E[U_1 V_1^*]E[U_2 V_2^*] + E[U_1 V_2^*]E[U_2 V_1^*].\end{aligned} \tag{62}$$